\documentclass[10pt,letterpaper]{article}
\usepackage{opex3}
\usepackage{color}
\usepackage{cite}

\begin{document}

\title{Efficient hyperconcentration of nonlocal multipartite entanglement via the cross-Kerr nonlinearity}

\author{Xi-Han Li$^{1,2*}$ and Shohini Ghose$^{2,3}$}

\address{$^1$Department of Physics, Chongqing University, Chongqing, China\\
$^2$Department of Physics and Computer Science, Wilfrid Laurier University, Waterloo, Canada\\
$^3$Institute for Quantum Computing, University of Waterloo, Canada}

\email{*xihanlicqu@gmail.com} 



\begin{abstract}
We propose two schemes for concentration of hyperentanglement of nonlocal multipartite states which are simultaneously entangled in the polarization and spatial modes. One scheme uses an auxiliary single-photon state prepared according to the parameters of the less-entangled states. The other scheme uses two less-entangled states with unknown parameters to distill the maximal hyperentanglement. The procrustean concentration is realized by two parity check measurements in both the two degrees of freedom. Nondestructive quantum nondemolition detectors based on cross-Kerr nonlinearity are used to implement the parity check, which makes the unsuccessful instances reusable in the next concentration round. The success probabilities in both schemes can be made to approach unity by iteration. Moreover, in both schemes only one of the $N$ parties has to perform the parity check measurements. Our schemes are efficient and useful for quantum information processing involving
hyperentanglement.
\end{abstract}




\section{Introduction}
In the  past thirty years, many quantum information processing tasks such as quantum key distribution \cite{qkd1,qkd2,qkd3}, quantum secure direct communication \cite{qsdc1,qsdc2,qsdc3}, quantum teleportation \cite{tele}, quantum dense coding \cite{dense} and quantum repeaters \cite{repeater} have been proposed and realized. These tasks rely on a unique phenomenon of quantum world: entanglement. Entangled photon pairs are used as a quantum channel in long distance quantum communication. Typically, maximal entanglement is desirable in the subsequent quantum information processing in order to ensure security and efficiency. However, entanglement can only be prepared locally and it is fragile during transmission and storage. In other words, the degree of entanglement and the purity of quantum entangled states degrade due to the inevitable interaction with the environment, and that subsequently influences the security and efficiency of the quantum communication schemes.

There are several effective methods to obtain the desired maximally entangled states via noisy channels, one of which is called entanglement concentration. Entanglement concentration can distill maximal entanglement from an ensemble of less-entangled pure states. Pioneering work on this topic based on the Schmidt projection method was done in 1996  \cite{concen1}. Later, a practical entanglement concentration scheme based on linear optics was proposed and demonstrated \cite{concen_pbs1, concen_pbs2}. In 2008, Sheng et al. presented a nonlocal entanglement concentration scheme that exploited cross-Kerr nonlinearity, and had a higher efficiency and yield than those with linear optical elements \cite{concen_sheng1}. In 2011, Wang et al. also proposed a polarization-entanglement concentration scheme using the cross-Kerr nonlinearity, in which the efficiency of the quantum nondemolition detector was effectively improved by the double cross-phase modulation method \cite{concen_wang} . All these entanglement concentration schemes  succeed probabilistically. These schemes start with two identical less-entangled states to distill one maximally entangled state since the parameters of the low quality states are unknown. If the parameters are known, according to which additional states can be prepared and optical elements can be manipulated, only one less-entangled state is required to accomplish the concentration \cite{concen_deng, concen_sheng4}. Some other interesting entanglement concentration schemes have been proposed in the past decade \cite{concen_swap1, concen_swap2, concen_sheng2, concen_w, concen_w5, concen_w6, concen_w7}.

The photon, which is a common candidate for quantum communication, has many degrees of freedom (DOFs) to carry quantum information. States which are entangled simultaneously in more than one DOF are called hyperentangled states \cite{hyper1,hyper2,hyper3}. They can be prepared with current technology. Since each photon carries more than one qubit  of information, hyperentanglement can increase the capacity of quantum information processing \cite{dense1,computation, hyper4}. Hyperentanglement has been proposed as a resource for many important applications in quantum information processing, such as quantum key distribution\cite{qkd}, entanglement purification \cite{puri_pdc,puri1,depp,odepp,odepp1,omdepp} and state analysis \cite{bsa1,bsa2,bsa3,bsa4,bsa6,gsa}. Hyperentanglement is a very promising quantum resource in quantum information processing.
However, hyperentangled states cannot escape from channel noise, which degrades the degree of entanglement in each DOF. Recently, the question of concentration of hyperentanglement has attracted much attention. In 2013, Ren et al. presented two hyperentanglement concentration schemes for a two-photon four-qubit system based on linear optical elements
\cite{hc1}. Later, we discussed the hyperconcentration of $N$-photon hyperentangled states via linear optics  \cite{hc10}. Although linear optics schemes are experimentally more feasible using current technology, the success probability of these schemes is restricted by the degree of entanglement of the less-entangled state to be concentrated. In 2013, one of us also proposed two hyperconcentration protocols for two-photon states with known and unknown parameters, respectively \cite{hc2}. In these schemes, both linear optics and nonlinearities are employed and some failed instances can be reused in the next round. Thus the success probabilities are higher than the schemes that only resort to linear optics. In the same year, Chen et al. proposed another hyperconcentration scheme for two-photon states based on projection measurements \cite{hc3}. Their success probability can be improved greatly by iteration. Moreover, Ren et al. proposed hyperentanglement concentration protocols assisted by diamond NV centers inside photonic crystal cavities \cite{hc4} and by quantum-dot spins inside
optical microcavities \cite{hc5}, respectively. Both these schemes rely on the interaction between photons and other systems to realize the parity check, which is the key step in entanglement concentration schemes.

In this letter, we present two hyperconcentration schemes for multipartite hyperentangled systems, that use the cross-Kerr nonlinearity. One scheme uses an additional single-photon two-qubit state prepared according to the parameters of the less-entangled state. The other uses two less-entangled states with unknown parameters. Both these protocols involve two parity check measurements on both two DOFs, i.e., the polarization state and the spatial mode. The parity check is realized via nondestructive quantum demolition detection established by the cross-Kerr nonlinearity.  In both these schemes, only one party has to perform the parity check measurements, while the other $N-1$ parties do nothing or just simple single-photon measurements. The unsuccessful instances in each round can be reconcentrated  to achieve higher success probabilities in both schemes. All these features make our schemes efficient and useful for long distance quantum communication.

\section{Quantum nondemolition detections}
Before we describe our hyperconcentration schemes, we introduce the principle of cross-Kerr nonlinearity and quantum nondemolition detections based on the weak cross-Kerr effect. The cross-Kerr nonlinearity is an interaction between a signal state $\vert \psi \rangle_s$ and a coherent probe beam $\vert \alpha \rangle_p$ in the nonlinear medium. The Hamiltonian of a cross-Kerr nonlinearity is \cite{kerr1,kerr2}
\begin{eqnarray}
H=\hbar\chi a_s^{\dag}a_s a_p^{\dag}a_p.
\end{eqnarray}
Here $a_s^{\dag}(a_p^{\dag})$ and $a_s(a_p)$ are the creation operator and destruction operator for the signal (probe) state, respectively. The coupling strength of the nonlinearity is given by $\chi$. The basic principle is that after the interaction with the signal state, the coherent probe beam picks up a phase shift depending on the photon number $n$ of the signal state,
\begin{eqnarray}
\vert \alpha \rangle_p \rightarrow  \vert \alpha e^{in\theta}\rangle,
\end{eqnarray}
where $\theta=\chi t$ and $t$ is the interaction time which can be controlled. Then by reading the phase shift via an $x$-quadrature measurement, the number of the photons can be measured without destroying the photon state. The cross-Kerr nonlinearity have been widely used in quantum information processing in the past years \cite{kerr2, concen_sheng1, concen_sheng4, concen_w5,hc2,hc3}.

\begin{center}
\begin{figure}[!h]
\centering\includegraphics*[height=1.5in]{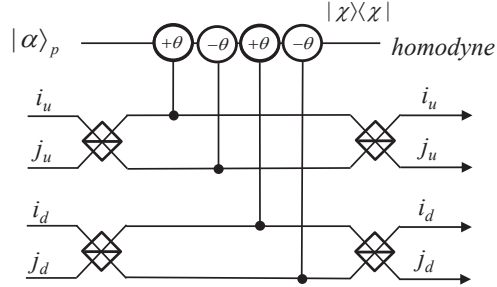}
\caption{The schematic diagram of the polarization parity check device based on the cross-Kerr nonlinearity. Polarizing beam splitters (PBSs) are used to transmit the horizontal polarization $\vert H \rangle$ and reflect the vertical polarization $\vert V \rangle$. $\pm \theta=\chi t$ represent the cross-Kerr nonlinear media that introduce the phase shift $\theta$ when there is a photon passing through the media. $\vert \chi \rangle\langle \chi \vert$ is the homodyne measurement that can distinguish different phase shifts. }
\end{figure}
\end{center}

The polarization parity check (PPC) device based on the cross-Kerr nonlinearity is shown in Fig. 1. The $i$ and $j$ denote two input modes for the two photons and $u$ and $d$ represent the two spatial modes "up" and "down" of each particle. The polarizing beam splitter (PBS) transmits  horizontal polarization states $\vert H \rangle$ and reflects the vertical ones $\vert V\rangle$. After interactions with the coherent probe beam, photons are guided back to their paths. This device can measure the parity of the polarization state without destroying the photons. The even-parity state $\vert HH\rangle$ and $\vert VV\rangle$ results in no phase shift of the coherent state while the odd-parity states $\vert HV \rangle$ and $\vert VH\rangle$ cause phase shift $+\theta$ and $-\theta$, respectively. Since the states $\vert \alpha e^{\pm i\theta}\rangle$ cannot be discriminated by the $x$-quadrature homodyne measurement, these two odd-parity states cannot be distinguished by this setup. This device checks the parity of the polarization state without disturbing the spatial mode DOF.

We can also design a spatial mode parity check (SPC) setup utilizing the cross-Kerr effect, as shown in Fig. 2. When the two photons have the same spatial mode, i.e., $\vert uu\rangle$ or $\vert dd\rangle$, there is no phase shift on the coherent state. Otherwise, a phase shift $\pm \theta$ can be detected. This device is used to check the parity of the spatial mode DOF without disturbing the polarization states.

\begin{center}
\begin{figure}[!h]
\centering\includegraphics*[height=1.5in]{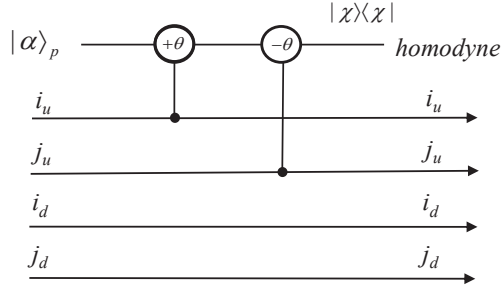}
\caption{The schematic diagram of the spatial mode parity check device that is used to check the parity of the spatial modes of photons $i$ and $j$.  $\pm \theta$ represent the cross-Kerr nonlinear media which introduce the phase shift $\theta$ when there is a photon passing through the media. $\vert \chi \rangle\langle \chi \vert$ is the homodyne measurement that can discriminate different phase shifts. }
\end{figure}
\end{center}

With these two nondemolition parity check devices, we can concentrate partially hyperentangled $N$-photon GHZ states that are simultaneously entangled in the polarization and the spatial mode.

\section{Hyperconcentration}

The partially hyperentangled $N$-photon GHZ state can be written as
\begin{eqnarray}
\vert \Psi\rangle_{AB...C}=(\alpha \vert HH...H\rangle +\beta\vert VV...V\rangle)
\otimes(\delta\vert a_{u}b_{u}...c_{u}\rangle+\eta \vert a_{d}b_{d}...c_{d}\rangle).
\end{eqnarray}
The parameters satisfy the normalization conditions $\vert \alpha\vert ^2+\vert \beta\vert ^2=\vert \delta\vert^2+\vert \eta\vert^2=1$. The subscripts $A$, $B$,...,$C$ represent the particles belong to the $N$ nonlocal parties Alice, Bob,...,Charlie and $x_u$ and $x_d$ are the two potential spatial modes of particle $X (X=A,B,...C)$.  The purpose of the hyperconcentration scheme is to get a maximally hyperentangled state GHZ state as follows.
\begin{eqnarray}
\vert \Phi\rangle_{AB...C}=\frac{1}{\sqrt{2}}(\vert HH...H\rangle +\vert VV...V\rangle)\otimes\frac{1}{\sqrt{2}}(\vert a_{u}b_{u}...c_{u}\rangle+\vert a_{d}b_{d}...c_{d}\rangle).
\end{eqnarray}
To distill the maximal hyperentangled state from the less-entangled samples, two hyperconcentration schemes are presented. The first scheme utilizes an additional single-photon state, with the precondition that the parameters of the less-entangled states are known. The second one exploits two less-entangled states for which the parameters are unknown.

\subsection{Hyperconcentration assisted by additional photon}
In this scheme, the parameters of the initial states are assumed to be known, according to which one party Alice can prepare an additional state
\begin{eqnarray}
\vert \varphi \rangle_{X}=(\alpha\vert V \rangle +\beta\vert H \rangle)_{X}\otimes(\delta\vert x_d \rangle+ \eta \vert x_u \rangle)_{X}.
\end{eqnarray}
Here $x_u$ and $x_d$ are the two spatial modes of the photon $X$. The quantum state of the whole system composed of $N+1$ photons can be written as
\begin{eqnarray}
\vert \Psi' \rangle_{AB...CX}&=&[\alpha^2\vert HH...HV\rangle+\beta^2\vert VV...VH\rangle\nonumber\\&&+\alpha\beta(\vert HH...HH \rangle+\vert VV...VV\rangle)]\nonumber\\&&\otimes [\delta^2\vert a_{u}b_{u}...c_{u}x_{d}\rangle
+\eta^2\vert a_{d}b_{d}...c_{d}x_{u}\rangle\nonumber\\&&+\delta\eta(\vert a_{u}b_{u}...c_{u}x_{u}\rangle+\vert a_{d}b_{d}...c_{d}x_{d}\rangle)].
\end{eqnarray}
\begin{center}
\begin{figure}[!h]
\centering\includegraphics*[width=2.5in]{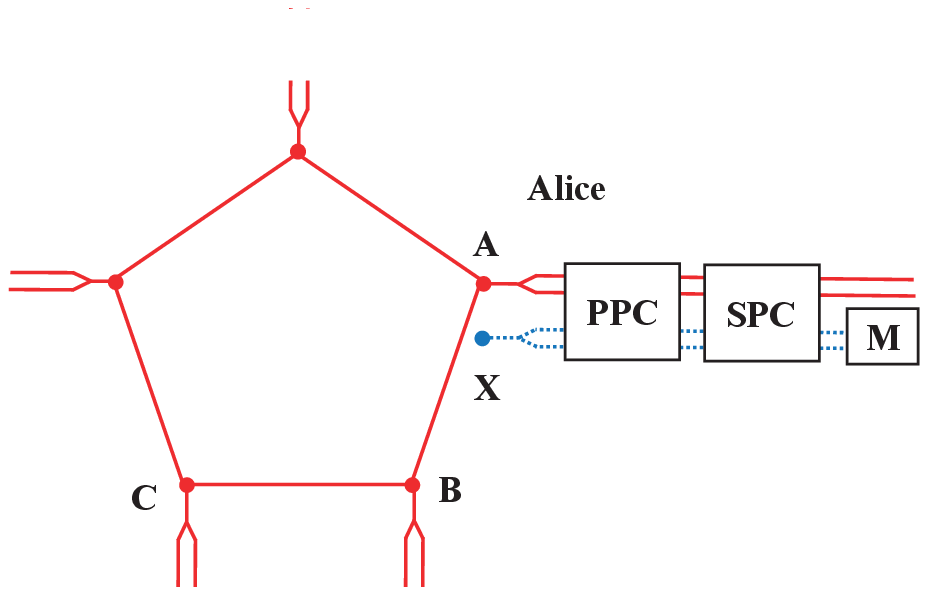}
\caption{Schematic diagram of the proposed hyperconcentration protocol assisted by an auxiliary photon. An additional photon $X$ is prepared by Alice based on the parameters of the original state. Then the PPC and SPC are performed on Alice's two photons $A$ and $X$ and the particle $X$ is measured in the diagonal basis of both DOFs, denoted by M. By selecting only even-parity outcomes in both the DOFs and then measuring the additional photon in the diagonal basis, the remote $N$ parities share the maximally hyperentangled GHZ state with a certain probability.}
\end{figure}
\end{center}
As shown in Fig. 3, Alice performs the polarization parity check and spatial mode parity check on photons $A$ and $X$ in sequence. She only selects the situation where both photons are in the even-parity state in both parity checks. Then the state becomes
\begin{eqnarray}
\vert \Psi' \rangle_{AB...CX}=\alpha\beta\delta\eta(\vert HH...HH \rangle+\vert VV...VV\rangle)]\otimes(\vert a_{u}b_{u}...c_{u}x_{u}\rangle+\vert a_{d}b_{d}...c_{d}x_{d}\rangle).
\end{eqnarray}
This is the maximally hyperentangled $N+1$-photon GHZ state. Then Alice measures the auxiliary photon $X$ in the diagonal basis of both DOFs $\vert \pm_p\rangle=\frac{1}{\sqrt{2}}(\vert H\rangle\pm\vert V\rangle)(\vert \pm_s\rangle=\frac{1}{\sqrt{2}}(\vert x_u\rangle\pm\vert x_d\rangle))$. The setup for the single-photon two-qubit measurement is shown in Fig. 4.
\begin{center}
\begin{figure}[!h]
\centering\includegraphics*[height=1.6in]{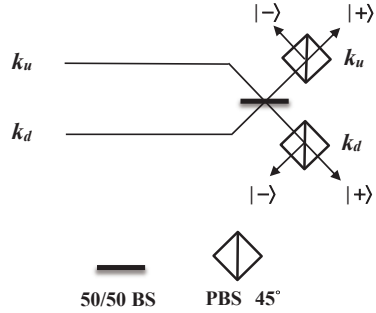}
\caption{Schematic diagram of the single-photon two-qubit measurement. Both the two DOFs of the single photon are measured in the diagonal basis. The measurement of spatial mode is realized by the beam splitter (BS) and of the polarization state is implemented by PBS at $45^\circ$. The effect of the balanced BS is $\vert u\rangle\rightarrow \frac{1}{\sqrt{2}}(\vert u\rangle+ \vert d\rangle)$, $\vert d\rangle\rightarrow \frac{1}{\sqrt{2}}(\vert u\rangle- \vert d\rangle)$.}
\end{figure}
\end{center}
There are four potential measurement results for photon $X$, according to which the state of the remaining  $N$ photons collapses to one of the states
\begin{eqnarray}
\vert \Psi' \rangle_{AB...C}=\alpha\beta\delta\eta(\vert HH...H \rangle\pm\vert VV...V\rangle)]\otimes(\vert a_{u}b_{u}...c_{u}\rangle\pm\vert a_{d}b_{d}...c_{d}\rangle).
\end{eqnarray}
If the measurement result is $\vert +\rangle$ ($\vert -\rangle$) for the polarization/spatial mode DOF, a unitary operation $I$ ($\sigma_z$) should be performed on the polarization/spatial mode of the photon $A$ to get $\vert \Phi\rangle_{AB...C}$. Here $\sigma^p_z=\vert H\rangle\langle H\vert-\vert V\rangle\langle V\vert$ and $\sigma^s_z=\vert a_u\rangle\langle a_u\vert-\vert a_d\rangle\langle a_d\vert$. The success probability of this $N$-photon  hyperconcentration scheme is $4\vert \alpha\beta\delta\eta\vert^2$, which is the same as that of the hyperconcentration scheme for a two-photon state via linear optics \cite{hc1}.

\subsection{Hyperconcentration using two less-entangled states}
If the parameters of the initial state are unknown, two identical less-entangled states are required to distill a maximal hyperentangled state probabilistically. The principle is shown in Fig. 5. Here we have two states $\vert \Psi\rangle_{A_1B_1...C_1}$ and $\vert \Psi\rangle_{A_2B_2...C_2}$.
\begin{center}
\begin{figure}[!h]
\centering\includegraphics*[width=2.8in]{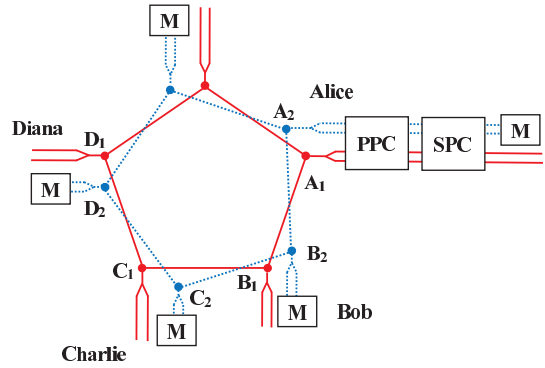}
\caption{Schematic diagram of the proposed hyperconcentration protocol which distills maximal hyperentanglement from two identical less-entangled nonlocal $N$-photon GHZ states (solid and dotted). Alice performs a PPC and SPC on her photons $A_1$ and $A_2$ to check the parity of the polarization and spatial mode DOFs and then measures $A_2$. The other $N-1$ parties (Bob, Charlie, etc) perform single-photon two-qubit measurements on their second photons. By selecting on the even-parity results of Alice's two parity checks, the distant $N$ parties share the maximally hyperentangled state probabilistically.}
\end{figure}
\end{center}

Before concentration, we transform the second state into
\begin{eqnarray}
\vert \Psi\rangle_{A_2B_2...C_2}=(\alpha \vert VV...V\rangle +\beta\vert HH...H\rangle)\otimes(\delta\vert a_{2d}b_{2d}...c_{2d}\rangle+\eta \vert a_{2u}b_{2u}...c_{2u}\rangle).
\end{eqnarray}
The flip of polarization can be realized by a half wave plate oriented at $45^\circ$. And the flip of spatial mode can be simply realized by changing their labels. These operations are not shown in Fig. 5 for simplicity. Then the state of the system composed of $2N$ photons is
\begin{eqnarray}
&&\vert \Xi\rangle_{A_1B_1...C_1A_2B_2...C_2}\nonumber\\&=&[\alpha^2\vert HH...HVV...V\rangle+\beta^2\vert VV...VHH...H\rangle\nonumber\\&&+\alpha\beta(\vert HH...HHH...H \rangle+\vert VV...VVV...V\rangle)]\nonumber\\&&\otimes [\delta^2\vert a_{1u}b_{1u}...c_{1u}a_{2d}b_{2d}...c_{2d}\rangle
+\eta^2\vert a_{1d}b_{1d}...c_{1d}a_{2u}b_{2u}...c_{2u}\rangle\nonumber\\&&+\delta\eta(\vert a_{1u}b_{1u}...c_{1u}a_{2u}b_{2u}...c_{2u}\rangle+\vert a_{1d}b_{1d}...c_{1d}a_{2d}b_{2d}...c_{2d}\rangle)].
\end{eqnarray}
Firstly, Alice inputs her qubits $A_1$ and $A_2$ into the polarization parity check device and then the spatial mode parity check device. She selects the cases in which both the parity check measurements give even parity results. The selected state can be written as
\begin{eqnarray}
\vert \Xi'\rangle_{A_1B_1...C_1A_2B_2...C_2}&=&\alpha\beta\delta\eta(\vert HH...HHH...H \rangle+\vert VV...VVV...V\rangle)\nonumber\\&&\otimes (\vert a_{1u}b_{1u}...c_{1u}a_{2u}b_{2u}...c_{2u}\rangle+\vert a_{1d}b_{1d}...c_{1d}a_{2d}b_{2d}...c_{2d}\rangle).
\end{eqnarray}
Then the $N$ parties measure their second photons with the single-photon two-qubit measurement setups. The state of the system composed of the remaining $N$ photons is
\begin{eqnarray}
\vert \Xi''\rangle_{A_1B_1...C_1}=\frac{1}{2}(\vert HH...H \rangle+(-1)^P\vert VV...V\rangle)\otimes (\vert a_{1u}b_{1u}...c_{1u}\rangle+(-1)^Q\vert a_{1d}b_{1d}...c_{1d}\rangle).
\end{eqnarray}
Here $P$ and $Q$ depend on all the $N$ parties' measurement outcomes. If the number of $\vert -_p\rangle$ is even (odd), $P=0$ $(1)$. And when the number of $d$ is even (odd), $Q=0$ $(1)$. Then one party can perform the corresponding phase-flip operation on his/her photon to obtain the desired state $\vert \Phi\rangle_{A_1B_1...C_1}$. The total success probability is again $4\vert \alpha\beta\delta\eta\vert ^2$.

\section{Improving the success probability by iteration}
In the last section we demonstrated the basic principle of our hyperconcentration schemes. The success of the hyperconcentration schemes is based on the two parity checks. When two even-parity outcomes occur, the concentration schemes succeed with probability $P_{ee}^{(1)}=4\vert \alpha\beta\delta\eta\vert^2$. Otherwise, these schemes fail. The total success probability is limited by the initial states, which is the same restriction on hyperconcentration schemes based on linear optics.

However, since we utilize quantum nondemolition detections  where the photons are intact after the measurement, the failed instances can be reused in the next round. Besides the even-parity results $``ee"$ in the both these two parity check measurements, there are three other possible outcomes, i.e., even-parity in polarization with odd-parity in spatial mode, odd-parity in spatial mode and even-parity in spatial mode and odd-parity in both DOFs. In these three cases, the state shared by the $N$ parties after the first concentration round can be changed to the following states, no matter which hyperconcentration method we use.
\begin{eqnarray}
\vert \Psi'_{eo}\rangle_{AB...C}&=&\alpha\beta(\vert HH...H\rangle+\vert VV...V\rangle)\otimes (\delta^2\vert  a_u b_u ...c_u\rangle +\eta^2\vert a_d b_d ...c_d\rangle), \\
\vert \Psi'_{oe}\rangle_{AB...C}&=&(\alpha^2\vert HH...H\rangle+\beta^2\vert VV...V\rangle)\otimes \delta\eta(\vert  a_u b_u ...c_u\rangle +\vert a_d b_d ...c_d\rangle), \\
\vert \Psi'_{oo}\rangle_{AB...C}&=&(\alpha^2\vert HH...H\rangle+\beta^2\vert VV...V\rangle)\otimes(\delta^2\vert  a_u b_u ...c_u\rangle +\eta^2\vert a_d b_d ...c_d\rangle).
\end{eqnarray}
The subscripts $``eo"$, $``oe"$ and $``oo"$ indicate the parity check results for the PPC and SPC, with $``e"$ being even and $``o"$ being odd. These three states are unnormalized. The probabilities of getting these three states are
\begin{eqnarray}
P^{(1)}_{eo}&=&2\vert \alpha\vert ^2\vert \beta\vert^2(\vert\delta\vert^4+\vert\eta\vert^4),\\
P^{(1)}_{oe}&=&2\vert \delta\vert ^2\vert \eta\vert^2(\vert\alpha\vert^4+\vert\beta\vert^4),\\
P^{(1)}_{oo}&=&(\vert\alpha\vert^4+\vert\beta\vert^4)(\vert\delta\vert^4+\vert\eta\vert^4).
\end{eqnarray}
The superscript $``(1)"$ denotes the first round of concentration.
These three less-entangled pure states have different coefficients, which can be written in a unified expression
\begin{eqnarray}
\vert \Psi^{(2)} \rangle_{AB...C}=(\alpha_i^{(2)}\vert HH...H\rangle+\beta_i^{(2)}\vert VV...V\rangle)\otimes (\delta_i^{(2)}\vert  a_u b_u ...c_u\rangle +\eta_i^{(2)}\vert a_d b_d ...c_d\rangle).
\end{eqnarray}
Here $(i=eo,oe,oo)$ and the normalized coefficients are shown in Table I.

\begin{table}[!h]
\centering
\caption{The normalized coefficients for the three less-entangled states corresponding to the failed instances of the parity checks.}
\begin{tabular}{cc|c|c|c|c}
\hline
& $State$ & $\alpha$ & $\beta$  & $\delta$ & $\eta$\\\hline

& $\vert \Psi^{(2)}_{eo} \rangle_{AB...C}$  &  $\alpha^{(2)}_{eo}=\frac{1}{\sqrt{2}}$ & $\beta^{(2)}_{eo}=\frac{1}{\sqrt{2}}$&$\delta^{(2)}_{eo}=\frac{\delta^2}{\sqrt{\delta^4+\eta^4}}$ & $\eta^{(2)}_{eo}=\frac{\eta^2}{\sqrt{\delta^4+\eta^4}}$ \\\hline

& $\vert \Psi^{(2)}_{oe} \rangle_{AB...C}$  &  $\alpha^{(2)}_{oe}=\frac{\alpha^2}{\sqrt{\alpha^4+\beta^4}}$ & $\beta^{(2)}_{oe}=\frac{\beta^2}{\sqrt{\alpha^4+\beta^4}}$ &  $\delta^{(2)}_{oe}=\frac{1}{\sqrt{2}}$ & $\eta^{(2)}_{oe}=\frac{1}{\sqrt{2}}$  \\\hline

& $\vert \Psi^{(2)}_{oo} \rangle_{AB...C}$  &  $\alpha^{(2)}_{oo}=\frac{\alpha^2}{\sqrt{\alpha^4+\beta^4}}$ & $\beta^{(2)}_{oo}=\frac{\beta^2}{\sqrt{\alpha^4+\beta^4}}$ &  $\delta^{(2)}_{oo}=\frac{\delta^2}{\sqrt{\delta^4+\eta^4}}$ &  $\eta^{(2)}_{oo}=\frac{\eta^2}{\sqrt{\delta^4+\eta^4}}$  \\\hline

\hline
\end{tabular}\label{Table1}

\end{table}

For each of the states in Eq. (19), we can use our hyperconcentration methods to further concentrate them. On one hand, an additional photon prepared in state $\vert \varphi \rangle_{X}=(\alpha_i^{(2)}\vert V\rangle+\beta_i^{(2)}\vert H\rangle)\otimes (\delta_i^{(2)}\vert  x_d\rangle +\eta_i^{(2)}\vert x_{u}\rangle)$ can be used to assist the concentration of the corresponding state. On the other hand, two less-entangled states with identical coefficients can be used to distill the maximally hyperentangled states in the next round.

In the second round, from Eqs (13) and (14), it is clear that for the state $\vert \Psi^{(2)}_{eo} \rangle_{AB...C}$ $(\vert \Psi^{(2)}_{oe} \rangle_{AB...C})$, the polarization (spatial mode) state is already in the desired form. Hence both even and odd outcomes of the polarization (spatial mode) parity check results ``$ee$" and ``$oe (eo)$" result in our desired maximally hyperentangled state. However, for the $\vert \Psi^{(2)}_{oo} \rangle_{AB...C}$ state, only the even-parity outcome $``ee"$ results in the target state. Therefore, the probability of obtaining the desired state in the second round is
\begin{eqnarray}
P^{(2)}=P_{eo}^{(1)}(P_{eo-ee}^{(2)}+P_{eo-oe}^{(2)})+P_{oe}^{(1)}(P_{oe-ee}^{(2)}+P_{oe-eo}^{(2)})
+P_{oo}^{(1)}P_{oo-ee}^{(2)}
\end{eqnarray}
Here $P^{(k)}_{i-j}$ represent the probability of obtaining the $j$ $(j=ee,eo,oe,oo)$ result in the $k$th round from the state $\vert \Psi_{i}^{(k)}\rangle$ $(i=eo,oe,oo)$.

Given the analysis of the second round, we can see that in the $k$th round, for the state $\vert \Psi^{(k)}_{eo} \rangle_{AB...C}$ $(\vert \Psi^{(k)}_{oe} \rangle_{AB...C})$, the polarization (spatial mode) state is already in the desired form.  Thus we can calculate that in the $k$th round, the probabilities of success (failure) of getting the desired state from the $\vert \Psi^{(k)}_{eo} \rangle_{AB...C}$ and $\vert \Psi^{(k)}_{oe} \rangle_{AB...C}$ which are generated from the $(k-1)$th round are
\begin{eqnarray}
P^{(k)}_{eo,s}&=&P^{(k)}_{eo-ee}+P^{(k)}_{eo-oe}=\frac{2\vert \delta\eta \vert^{2^k}}{(\vert \delta\vert^{2^k}+\vert \eta\vert^{2^k})^2},\\
P^{(k)}_{eo,f}&=&P^{(k)}_{eo-eo}+P^{(k)}_{eo-oo}=\frac{\vert \delta\vert^{2^{k+1}}+\vert \eta\vert^{2^{k+1}}}{(\vert \delta\vert^{2^k}+\vert \eta\vert^{2^k})^2},\\
P^{(k)}_{oe,s}&=&P^{(k)}_{oe-ee}+P^{(k)}_{oe-eo}=\frac{2\vert \alpha\beta \vert^{2^k}}{(\vert \alpha\vert^{2^k}+\vert \beta\vert^{2^k})^2},\\
P^{(k)}_{oe,f}&=&P^{(k)}_{oe-oe}+P^{(k)}_{oe-oo}=\frac{\vert \alpha\vert^{2^{k+1}}+\vert \beta\vert^{2^{k+1}}}{(\vert \alpha\vert^{2^k}+\vert \beta\vert^{2^k})^2}.
\end{eqnarray}
Here $P^{(k)}_{i,s}$ and $P^{(k)}_{i,f}$ $(i=eo,oe)$ denote the success and failure probabilities for obtaining the maximally hyperentangled state from $\vert \Psi^{(k)}_{i} \rangle_{AB...C}$ in the $k$th round.

For the $\vert \Psi^{(k)}_{oo} \rangle_{AB...C}$ state which corresponds to two odd-parity results in the $(k-1)$th round, the probabilities of the four parity check results are
\begin{eqnarray}
P^{(k)}_{oo-ee}&=&\frac{4\vert \alpha\beta\delta\eta \vert^{2^k}}{(\vert \alpha\vert^{2^k}+\vert \beta\vert^{2^k})^2(\vert \delta\vert^{2^k}+\vert \eta\vert^{2^k})^2}=P^{(k)}_{eo,s}P^{(k)}_{oe,s}, \\
P^{(k)}_{oo-eo}&=&\frac{2\vert \alpha\beta \vert^{2^k}(\vert \delta\vert^{2^{k+1}}+\vert \eta\vert^{2^{k+1}})}{(\vert \alpha\vert^{2^k}+\vert \beta\vert^{2^k})^2(\vert \delta\vert^{2^k}+\vert \eta\vert^{2^k})^2}=P^{(k)}_{eo,f}P^{(k)}_{oe,s},\\
P^{(k)}_{oo-oe}&=&\frac{2\vert \delta\eta\vert^{2^k}(\vert \alpha\vert^{2^{k+1}}+\vert \beta\vert^{2^{k+1}})}{(\vert \alpha\vert^{2^k}+\vert \beta\vert^{2^k})^2(\vert \delta\vert^{2^k}+\vert \eta\vert^{2^k})^2}=P^{(k)}_{eo,s}P^{(k)}_{oe,f},\\
P^{(k)}_{oo-oo}&=&\frac{(\vert \alpha\vert^{2^{k+1}}+\vert \beta\vert^{2^{k+1}})(\vert \delta\vert^{2^{k+1}}+\vert \eta\vert^{2^{k+1}})}{(\vert \alpha\vert^{2^k}+\vert \beta\vert^{2^k})^2(\vert \delta\vert^{2^k}+\vert \eta\vert^{2^k})^2}=P^{(k)}_{eo,f}P^{(k)}_{oe,f}.
\end{eqnarray}
\begin{center}
\begin{figure}[b]
\centering\includegraphics[width=3.5in]{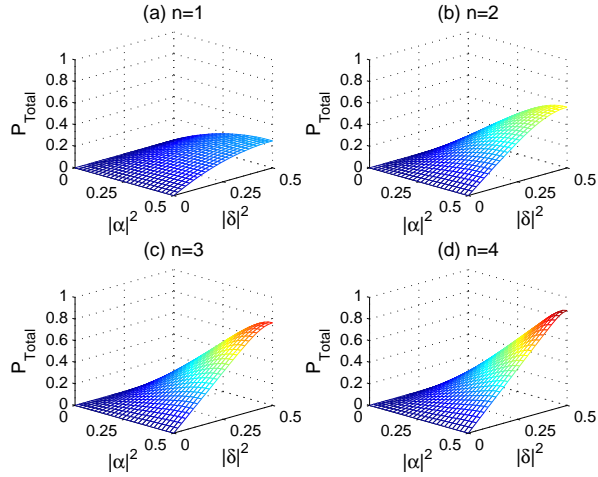}
\caption{The total success probability of obtaining the maximally hyperentangled state depends on the parameters of the initial state $\vert \alpha\vert^2$ and $\vert \delta\vert^2$. Different Figs. correspond to the schemes with different number of iterations.}
\end{figure}
\end{center}
Each round of concentration succeeds probabilistically and the failed instances can be reused in the next round. Then we can compute the success probability of the $k$th $(k>2)$ round,

\begin{eqnarray}
P^{(k)}&=&[P^{(1)}_{eo}+P^{(1)}_{oo}(P^{(2)}_{oe,s}+P^{(2)}_{oe,f}P^{(3)}_{oe,s}+...
P^{(2)}_{oe,f}P^{(3)}_{oe,f}...P^{(k-2)}_{oe,f}P^{(k-1)}_{oe,s})]P^{(2)}_{eo,f}...P^{(k-1)}_{eo,f}P^{(k)}_{eo,s}\nonumber\\
&&+[P^{(1)}_{oe}+P^{(1)}_{oo}(P^{(2)}_{eo,s}+P^{(2)}_{eo,f}P^{(3)}_{eo,s}+...
P^{(2)}_{eo,f}P^{(3)}_{eo,f}...P^{(k-2)}_{eo,f}P^{(k-1)}_{eo,s})]P^{(2)}_{oe,f}...P^{(k-1)}_{oe,f}P^{(k)}_{oe,s}\nonumber\\
&&+P^{(1)}_{oo}P^{(2)}_{eo,f}P^{(2)}_{oe,f}P^{(3)}_{eo,f}P^{(3)}_{oe,f}...P^{(k-1)}_{eo,f}P^{(k-1)}_{oe,f}P^{(k)}_{eo,s}
P^{(k)}_{oe,s}.
\end{eqnarray}

By iterating the hyperconcentration process $n$ times, the total success probability is
\begin{eqnarray}
P_{Total}=\sum^n_{k=1} P^{(k)},
\end{eqnarray}
which depends on the number of iteration and parameters of the initial state. The relation between the total success probability, the initial parameter and the iteration number is shown in Fig. 6. The four plots show the success probabilities for  $n=1,2,3,4$ iterations. We find the total success probability monotonically increase with the parameters $\vert \alpha \vert^2$ and $\vert \delta \vert^2$ in the range $[0,0.5]$, and there is an obvious improvement with more rounds. After $n=5$ iterations, the maximum of the success probability will be larger than $90\%$, which is much larger than that of the hyperconcentration scheme with linear optics \cite{hc1,hc10}.

\section{Discussion and Summary}
This paper presents two efficient schemes for concentration of nonlocal $N$-photon hyperentanglement, using the cross-Kerr nonlinearity. The first one utilizes an auxiliary photon prepared according to the known parameters of the initial state and the second one uses two identical less-entangled states with unknown parameters. In each round of the hyperconcentration schemes, only one party has to perform the polarization parity check and the spatial mode parity check. The other $(N-1)$ parties do nothing or just perform single-photon two-qubit measurements that can be realized with current technology. In both schemes, after the first hyperconcentration round, the distant $N$ parties will share the maximally hyperentangled state with a certain probability. The success probability can be greatly improved by implementing more rounds of concentration.

In our two schemes, the hyperconcentration is implemented based on two parity checks, one for the polarization state and the other for the spatial mode. A quantum nondemolition measurement that exploits the cross-Kerr effect was used to make the schemes iterable. Therefore, the practical efficiency of our schemes depends on the efficiency of the cross-Kerr nonlinearity. Although the natural cross-Kerr nonlinearities are weak and the Kerr phase shift is small at the single-photon level, recent research shows it is promising to use the effect in the near future. In 2011, He et al. investigated the interaction between a single photon and a coherent state and made the treatment of coherent state-single photon interactions
more realistic \cite{kerr3}. In the same year, it was shown that the amplification of a cross-Kerr phase shift to an observable value by using weak-value amplification is possible \cite{kerr4}. Moreover, giant Kerr nonlinearity of the probe and the
signal pulses was shown in Ref. \cite{kerr5}. Our schemes only require a small phase shift, as long as  it can be distinguished from zero. This feature makes our schemes more practical than others which need a giant Kerr nonlinearity. Moreover, other kinds of nonlinear interaction can also provide feasible ways to realize the parity check we need \cite{other1,other2,other3,other4,other5}.

In our schemes, the total success probability can be increased by iteration and is not limited by the coefficients of the initial state. This makes our schemes much more efficient than the schemes using linear optics. Our two hyperconcentration schemes are for two different situations, depending on whether the parameters of the initial state are known or not. In a practical setting, if the state parameters are unknown, the parties can either use our second scheme for the case of unknown parameters, or they can first perform parameter estimation and then use our first scheme for a state with known parameters.
The estimation of parameters consumes extra quantum resources. Since the form of the state is already known, Alice can just measure her photons in the computational basis ($\vert H\rangle/\vert V\rangle$ or $\vert a_u\rangle/\vert a_d\rangle$  ) in both the DOFs. If a sufficient number of samples is measured, the value of parameters can be deduced with the desired accuracy from the probabilities of the results for each DOF.
In our first scheme with known parameters, each concentration round consumes $N+1$ photons to distill the $N$-photon maximally entangled state while in the second scheme $2N$ photons are used. Thus depending on the resources required for parameter estimation and the size of $N$, the parties can decide whether it is better to use the first scheme together with parameter estimation, or the second scheme without parameter estimation. Although the success probabilities of these two schemes can be improved to nearly 100\% in principle, these two schemes can only distill maximal hyperentanglement from less-entangled pure states. The efficiencies of hyperentanglement sharing will also be affected by other kinds of noise and photon loss. If photon loss occurs, the states can be rejected by postselection, or photon loss can be addressed by introducing existing techniques such as the targeted method \cite{loss}. Other kinds of decoherence require further investigation.


To sum up, we have proposed two efficient hyperconcentration schemes for multipartite hyperentangled states that are entangled both in polarization and spatial mode. In both schemes only one party has to perform the parity check measurements. The success probability can be improved to close to 100\%  by iteration. Since hyperentanglement has many potential applications, our schemes may be useful for quantum information processing involving hyperentanglement.

\section*{Acknowledgments}
XL is supported by the National Natural Science
Foundation of China under Grant No. 11004258, the Fundamental Research Funds for the Central Universities under
Grant No.CQDXWL-2012-014 and the Natural Science Foundation Project of CQ CSTC 2011jjA90017. SG acknowledges support from the Ontario Ministry of Research and Innovation and the Natural Sciences and Engineering Research Council of Canada.
\end{document}